\input epsf.tex
 \documentstyle[12pt]{article}

\newcommand{\bmat}{\left(\begin{array}}
\newcommand{\emat}{\end{array}\right)}
\def\NPB{Nucl. Phys. B}

\def\yzero{\smash{\hbox{$y\kern-4pt\raise1pt\hbox{${}^\circ$}$}}}

\def\b{\beta}

\def\beq{\begin{equation}}
\def\eeq{\end{equation}}
\def\beqa{\begin{eqnarray}}
\def\eeqa{\end{eqnarray}}

\def\-{\hphantom{-}}

\def\s2{\frac{1}{\sqrt2}}

\def\beq{\begin{equation}}
\def\eeq{\end{equation}}
\def\beqa{\begin{eqnarray}}
\def\eeqa{\end{eqnarray}}

\def\IF{\relax{\rm I\kern-.18em F}}
\def\II{\relax{\rm I\kern-.18em I}}
\def\IP{\relax{\rm I\kern-.18em P}}
\def\IC{\relax\hbox{\kern.25em$\inbar\kern-.3em{\rm C}$}}
\def\IR{\relax{\rm I\kern-.18em R}}

\def\cp{{\cal P}}

\def\Dsl{\,\raise.15ex\hbox{/}\mkern-13.5mu D} 
\def\IZ{Z\kern-.4em  Z}

 \def\cp#1{\relax\ifmmode {\IP\kern-2pt{}_{#1}}\else $\IP\kern-2pt{}_{#1}$\=fi}

\catcode`\@=11
\newdimen\@rotdimen
\newbox\@rotbox

\def\@vspec#1{\special{ps:#1}}
\def\@rotstart#1{\@vspec{gsave currentpoint currentpoint translate
   #1 neg exch neg exch translate}}
\def\@rotfinish{\@vspec{currentpoint grestore moveto}}
\def\@rotr#1{\@rotdimen=\ht#1\advance\@rotdimen by\dp#1%
   \hbox to\@rotdimen{\hskip\ht#1\vbox to\wd#1{\@rotstart{90 rotate}%
   \box#1\vss}\hss}\@rotfinish}
\def\@rotl#1{\@rotdimen=\ht#1\advance\@rotdimen by\dp#1%
   \hbox to\@rotdimen{\vbox to\wd#1{\vskip\wd#1\@rotstart{270 rotate}%
   \box#1\vss}\hss}\@rotfinish}%
\def\@rotu#1{\@rotdimen=\ht#1\advance\@rotdimen by\dp#1%
   \hbox to\wd#1{\hskip\wd#1\vbox to\@rotdimen{\vskip\@rotdimen
   \@rotstart{-1 dup scale}\box#1\vss}\hss}\@rotfinish}%
\def\@rotf#1{\hbox to\wd#1{\hskip\wd#1\@rotstart{-1 1 scale}%
   \box#1\hss}\@rotfinish}%
\def\rotate{\@ifnextchar[{\@rotate}{\@rotate[l]}}
\def\@rotate[#1]#2{\setbox\@rotbox=\hbox{#2}\@nameuse{@rot#1}\@rotbox}

\catcode`\@=12

\begin{document}

\makeatletter
\@addtoreset{equation}{section} \makeatother
\renewcommand{\theequation}{\thesection.\arabic{equation}}
\pagestyle{empty}
\pagestyle{empty}
\rightline{FTUAM-02-10}
\rightline{IFT-UAM/CSIC-02-09}
\rightline{\today}
\vspace{0.5cm}
\setcounter{footnote}{0}

\begin{center}
{\LARGE{ New Standard Model Vacua from Intersecting Branes}}
\\[7mm]
{\Large{{  C. ~Kokorelis} }
\\[2mm]}
\small{ Dep/to de F\'\i sica Te\'orica C-XI and 
Instituto de F\'\i sica 
Te\'orica C-XVI}
,\\[-0.3em]
{ Universidad Aut\'onoma de Madrid, Cantoblanco, 28049, Madrid, Spain}
\end{center}
\vspace{3mm}


\begin{center}
{\small \bf ABSTRACT}
\end{center}
\begin{center}
\begin{minipage}[h]{14.5cm}
 We construct new D6-brane model vacua (non-supersymmetric) that have 
at low energy exactly
the
standard model spectrum (with right handed neutrinos).
The minimal version
of these models requires five stacks of branes.
and the construction is based on 
D6-branes
intersecting at angles in $D = 4$ type toroidal orientifolds of type
I strings.
Three $U(1)$'s become massive through their couplings to RR couplings
and from the two surviving anomaly free $U(1)$'s,
one is the
standard model 
hypercharge generator while the extra anomaly free
$U(1)$ could be broken from its non-zero couplings to RR
fields and also by
triggering a vev to 
previously massive particles. We suggest that extra
massless $U(1)$'s 
should be broken by requiring some intersection 
to respect $N=1$ supersymmetry thus supporting the
appearance of massless
charged singlets at the supersymmetric intersection.
Proton is stable as
baryon number is gauged  
and its anomalies are cancelled through a generalized Green-Schwarz
mechanism. 
Neutrinos are of Dirac type with small masses, as in the four
stack standard models
of hep-th/0105155,
as a result of the existence of a similar PQ like-symmetry.
The models are unique in the sense that they predict the existence
of only one supersymmetric
particle, the superpartner
of $\nu_R$.

\end{minipage}                 
\end{center}

\newpage
\setcounter{page}{1}
\pagestyle{plain}
\renewcommand{\thefootnote}{\arabic{footnote}}
\setcounter{footnote}{0}

\section{Introduction}

Over the years, one of the most difficult questions
that string theory is facing is the selection of the particular vacuum
that includes at low energy all the necessary
ingredients of the observable
standard model spectrum a low energies.
In the absence of an underlying principle of
picking up a particular
vacuum, one can search for the model with the correct low energy particle
content. Such attempts have by far been explored in the context
of heterotic string theory as well to branes at 
singularities \cite{iba, luis3}. The main 
characteristics of the models involved 
include the three generation massless spectrum of the standard model (SM)
accompanied by the presence of exotic matter and /or gauge group factors. 
However, recently there has been some progress 
in the study of string models as it has been possible in 
\cite{louis2} to
derive, at low energy,
just the SM  spectrum together with right handed neutrinos.
 The models were studied in the context of 
intersecting branes and 
have some satisfactory properties including proton stability and small 
neutrino (of Dirac type) masses.

 The purpose of this paper is 
to extend the four stack construction of type I four
dimensional toroidal
orientifolded six-torus construction of \cite{louis2}, 
to a different structure that
involves one extra $U(1)$ at the string scale and produces just the standard
model (SM) at low energies. Note that 
in \cite{louis2} one starts with a $U(3)_a \otimes 
U(2)_b \otimes U(1)_c \otimes U(1)_d$ open gauge group structure at string 
scale energies. Note that in our construction,  
as in \cite{louis2}, right 
handed neutrinos are present along with the SM particle context at 
low energies.  
Additional non-supersymmetric models along the same Type I backgrounds, 
which give at low energy the SM structure, 
have been constructed in \cite{kokos1}. In the latter case one starts
with a four dimensional type I background on an  
orientifolded six dimensional factorized torus, which at the string
scale includes \footnote{The string scale gauge group structure, that 
includes
four stacks of branes, that is       
an $U(4)_c \otimes 
U(2)_L \otimes U(2)_R \otimes U(1)$.}  
a Pati-Salam gauge group. The models incorporate a 
number of interesting properties, like proton stability and small 
neutrino masses.  

In this work, one starts with five stacks of branes, namely with an 
$U(3)_a \otimes U(2)_b \otimes U(1)_c \otimes U(1)_d  \otimes  
U(1)_e$ at the string scale and get at low energy only the SM with 
right handed neutrinos. The models also allow
different generations of leptons-neutrinos to be placed at different
intersections, 
that could have interesting
implications for the phenomenology of the models.
We also note that from the five string scale $U(1)$'s,
four couple to RR fields and one survives massless
at low energies.
The latter U(1) corresponds to the hypercharge of the SM. 
\newline
The models are non-supersymmetric and are build on a background
of
D6-branes intersecting each other at non-trivial angles 
\cite{bele}, in an orientifolded
factorized six-torus,
while O$_6$ orientifold planes are on top of D6-branes \cite{louis2}.
Note that the latter picture is just the T-dual of type I 
backgrounds of \cite{tessera}, \cite{tessera1} which
make use of D9 branes with background fluxes.
Also, we note that the studied backgrounds are T-dual to
models with magnetic
deformations \cite{carlo}.

Additional models on backgrounds, in the context of intersecting branes, 
have been discussed 
in \cite{tessera2, luis1, alda, blume, cim, bere, eve}.  
For an other proposal for realistic SM D-brane model building,
based not 
on a particular string construction, see \cite{antokt}.

The proposed models have \footnote{ The different classes of models 
of this work maintain essential 
characteristics of the classes of models 
discussed in \cite{louis2}, including proton stability,
and sizes of neutrino masses within experimental limits.} the following
distinctive features :
\begin{itemize}

\item   The model starts with a gauge group at the string
scale $U(3) \times
U(2) \times U(1) \times U(1) \times U(1)$. The use of Green-Schwarz
mechanism in the model renders three of the U(1)'s
massive while a
combination of the other two remaining anomaly free
$U(1)$'s one
makes the standard model
hypercharge, while the other one \footnote{In an orthogonal basis, this U(1) 
will be broken only by the vev of a scalar created by turning N=1 SYSY on an
inbtersection.} gets
broken by its non-zero coupling to RR fields and
by turning on $N=1$ SUSY at an intersection,
while keeping the rest of the
intersections non-supersymmetric.

\item  Neutrinos get a Dirac mass, as lepton number L is a gauged
symmetry,
   of the right order \footnote{The same mechanism
was employed in \cite{louis2} for the four stack D6 orientifold counterpart
of the present SM's.}
in consistency with the LSND neutrino oscillation 
experiments \cite{LSND} as a consequence of the existence of a PQ-like
symmetry related to chiral symmetry breaking.

\item  Proton is stable due to the fact that baryon number B is an 
unbroken gauged 
global symmetry surviving at low energies whose anomalies 
cancel through a generalized Green-Schwarz mechanism.

\end{itemize}

The paper is organized as follows. In section two we describe
the general characteristics of the new standard model vacua
with particular emphasis on
how to 
calculate the fermionic spectrum from intersecting branes as well providing 
the classification of 
multi-parameter solutions to the RR tadpole cancellation conditions.
In section 3 we examine the cancellation of $U(1)$
anomalies via a
generalized 
Green-Schwarz (GS) mechanism \cite{iru, sa, iru1}
examining the conditions
under which the hypercharge generator remains light at low energies. 
In section 4 we study the Higgs sector providing for the tachyonic
scalars that are used for the
electroweak symmetry breaking of the model. We also discuss
the breaking of the extra anomaly free, other than hypercharge, $U(1)$,
by turning on $N=1$ supersymmetry at
an intersection.
In section 5, we examine the Yukawa couplings 
and the smallness of neutrino masses.
Section 6 contains our conclusions.

\section{New SM vacua from intersecting branes}

In the present work, we are going to 
describe new type I compactification vacua, that have as their low energy 
limit just
the observable standard model interactions.
The proposed three generation
non-supersymmetric
standard models make use of five stacks of branes at the string scale.
They will originate
from D6-branes wrapping on 3-cycles of 
toroidal orientifolds of type IIA in four dimensions.
Important characteristic of all vacua coming from these type I 
constructions is the replication, at each intersection, 
of the massless fermion spectrum
by an equal number of massive particles in the same representations 
and with the same quantum numbers.\newline 

Next, we describe the construction of the standard model. It is based on  
type I string with D9-branes compactified on a six-dimensional
orientifolded torus $T^6$, 
where internal background 
gauge fluxes on the branes are turned on. 
If we perform a T-duality transformation on the $x^4$, $x^5$, $x^6$, 
directions the D9-branes with fluxes are translated into D6-branes 
intersecting at 
angles. Note that the branes are not parallel to the
orientifold planes. Furthermore,
we assume that the D$6_a$-branes are wrapping 1-cycles 
$(n^i_a, m^i_a)$ along each of the ith-$T^2$
torus of the factorized $T^6$ torus, namely 
$T^6 = T^2 \times T^2 \times T^2$.
That means that we allow our torus to wrap factorized 3-cycles, that can 
unwrap into products of three 1-cycles, one for each $T^2$.
We define the homology of the 3-cycles as
\beq
[\Pi_a] =\ \prod_{i=1}^3(n^i_a [a_i] + m^i_a[b_i])
\label{homo1}
\eeq
while we define the 3-cycle for the orientifold images as
\beq
[\Pi_{a^{\star}}] =\ \prod_{i=1}^3(n^i_a [a_i] - m^i_a[b_i])
\label{homo2}
\eeq

In order to build the SM model structure a low energies, we consider
five stacks of D6-branes giving rise to their world-volume to an initial
gauge group $U(3) \times U(2) \times U(1) \times U(1) \times U(1)$ or
$SU(3) \times SU(2) \times U(1)_a \times U(1)_b \times U(1)_c \times 
U(1)_d \times U(1)_e$
at the string
scale.
Also, we consider the addition of NS B-flux \cite{flux}, 
such that the tori involved are not orthogonal, thus
avoiding an even number of families \cite{tessera}, 
and leading to effective tilted wrapping numbers as, 
\beq
(n^i, m ={\tilde m}^i + {n^i}/2);\; n,\;{\tilde m}\;\in\;Z.
\label{na2}
\eeq
In this way we allow semi-integer values for the m-wrapping numbers.  
\newline
Because of the $\Omega {\cal R}$ symmetry, 
where $\Omega$ is the worldvolume 
parity and $\cal R$ is the reflection on the T-dualized coordinates,
\beq
T (\Omega {\cal R})T^{-1}= \Omega {\cal R},
\label{dual}
\eeq
 each D$6_a$-brane
1-cycle, must have its $\Omega {\cal R}$ image 
partner $(n^i_a, -m^i_a)$. 

Chiral fermions gets localized at the intersections between branes,
 by stretched open strings between
intersecting D6-branes \cite{bele}. 
Subsequently, the chiral spectrum of the model may be obtained 
by solving simultaneously 
the intersection 
constraints coming from the existence of the different sectors and
the RR
tadpole cancellation conditions. 
Note that in the models we examine in this work,
there are a number of different sectors, 
which should be taken into account when computing the chiral spectrum.
We denote the action of 
$\Omega R$ on a sector $a, b$, by $a^{\star}, b^{\star}$, respectively.
The possible sectors are:

\begin{itemize}
 
\item The $a b + b a$ sector: involves open strings stretching between the 
D$6_a$ and D$6_b$ branes. Under the $\Omega R$ symmetry 
this sector is mapped to its image, $a^{\star} b^{\star}
+ b^{\star} a^{\star}$ sector.
The number, $I_{ab}$, of chiral fermions in this sector, 
transforms in the bifundamental representation
$(N_a, {\bar N}_a)$ of $U(N_a) \times U(N_b)$, and reads
\beq
I_{ab} = [\Pi_a] \cdot [\Pi_b] = ( n_a^1 m_b^1 - m_a^1 n_b^1)( n_a^2 m_b^2 - m_a^2 n_b^2 )
(n_a^3 m_b^3 - m_a^3 n_b^3),
\label{ena3}
\eeq
 where $I_{ab}$
is the intersection number of the wrapped cycles. Note that with the sign of
 $I_{ab}$ intersection, we denote the chirality of 
the fermions, where $I_{ab} > 0$ denotes left handed fermions.
Also negative multiplicity denotes opposite chirality.

\item The $a b^{\star} + b^{\star} a$ sector :
It involves chiral fermions transforming into the $(N_a, N_b)$
representation with multiplicity given by
\beq
I_{ab^{\star}} =\  [\Pi_a] \cdot [\Pi_{b^{\star}}]     =\ -( n_a^1 m_b^1 + m_a^1 n_b^1)( n_a^2 m_b^2 + m_a^2 n_b^2 )
(n_a^3 m_b^3 + m_a^3 n_b^3).
\label{ena31}
\eeq
Under the $\Omega R$ symmetry it transforms to itself.

\item the $a a^{\star}$ sector : under the $\Omega R$ symmetry it 
transforms to itself. From this sector the invariant intersections
will give 8$m_a^1 m_a^2 m_a^3$ fermions in the antisymmetric representation
and the non-invariant intersections that come in pairs provide us with
4$ m_a^1 m_a^2 m_a^3 (n_a^1 n_a^2 n_a^3 -1)$ additional 
fermions in the symmetric and 
antisymmetric representation of the $U(N_a)$ gauge group. However as we 
explain later, these
sectors will be absent from our models.
\end{itemize}

Any vacuum derived from the previous intersection constraints  
is subject additionally to constraints coming from RR tadpole cancellation 
conditions \cite{tessera}. That demands cancellation of 
D6-branes charges \footnote{Taken together with their
orientifold images $(n_a^i, - m_a^i)$  wrapping
on three cycles of homology
class $[\Pi_{\alpha^{\prime}}]$.}, wrapping on three cycles with
homology $[\Pi_a]$ and O6-plane 7-form
charges wrapping on 3-cycles  with homology $[\Pi_{O_6}]$. 
Note that the RR tadpole cancellation conditions
in terms of cancellations of RR charges in homology are
\beq
\sum_a N_a [\Pi_a]+\sum_{\alpha^{\prime}} 
N_{\alpha^{\prime}}[\Pi_{\alpha^{\prime}}] - 32
[\Pi_{O_6}]=0.
\label{homology}
\eeq  
In explicit form, the RR tadpole conditions read
\beqa
\sum_a N_a n_a^1 n_a^2 n_a^3 =\ 16,\nonumber\\
\sum_a N_a m_a^1 m_a^2 n_a^3 =\ 0,\nonumber\\
\sum_a N_a m_a^1 n_a^2 m_a^3 =\ 0,\nonumber\\
\sum_a N_a n_a^1 m_a^2 m_a^3 =\ 0.
\label{na1}
\eeqa
That guarantees absence of non-abelian gauge anomalies.
In D-brane model building, by considering $a$ 
stacks of D-brane configurations with 
$N_a, a=1, \cdots, N$, parallel branes one gets the gauge group 
 $U(N_1) \times U(N_2) \times \cdots \times U(N_a)$. 
Each $U(N_i)$ factor will give rise to an $SU(N_i)$
charged under the associated $U(1_i)$ gauge group factor that appears in 
the decomposition $SU(N_a) \times U(1_a)$.
For the five stack model that we examine in this work, 
the complete accommodation, where all other intersections are vanishing,  
of the fermion structure can be seen
in table (\ref{spectrum8}).
We note a number of interesting comments :
\newline 
\begin{table}[htb] \footnotesize
\renewcommand{\arraystretch}{1.5}
\begin{center}
\begin{tabular}{|c|c|c|c|c|c|c|c|c|}
\hline
Matter Fields & & Intersection & $Q_a$ & $Q_b$ & $Q_c$ & $Q_d$ & $Q_e$& Y
\\\hline
 $Q_L$ &  $(3, 2)$ & $I_{ab}=1$ & $1$ & $-1$ & $0$ & $0$ & $0$& $1/6$ \\\hline
 $q_L$  &  $2(3, 2)$ & $I_{a b^{\ast}}=2$ &  
$1$ & $1$ & $0$ & $0$  & $0$ & $1/6$  \\\hline
 $U_R$ & $3({\bar 3}, 1)$ & $I_{ac} = -3$ & 
$-1$ & $0$ & $1$ & $0$ & $0$ & $-2/3$ \\\hline    
 $D_R$ &   $3({\bar 3}, 1)$  &  $I_{a c^{\ast}} = -3$ &  
$-1$ & $0$ & $-1$ & $0$ & $0$ & $1/3$ \\\hline    
$L$ &   $2(1, 2)$  &  $I_{bd} = -2$ &  
$0$ & $-1$ & $0$ & $1$ & $0$ & $-1/2$  \\\hline    
$l_L$ &   $(1, 2)$  &  $I_{b e} = -1$ &  
$0$ & $-1$ & $0$ & $0$ & $1$ & $-1/2$  \\\hline    
$N_R$ &   $2(1, 1)$  &  $I_{cd} = 2$ &  
$0$ & $0$ & $1$ & $-1$ & $0$ & $0$  \\\hline    
$E_R$ &   $2(1, 1)$  &  $I_{c d^{\ast}} = -2$ &  
$0$ & $0$ & $-1$ & $-1$ & $0$ & $1$   \\\hline
  $\nu_R$ &   $(1, 1)$  &  $I_{c e} = 1$ &  
$0$ & $0$ & $1$ & $0$ & $-1$ & $0$ \\\hline
$e_R$ &   $(1, 1)$  &  $I_{c e^{\ast}} = -1$ &  
$0$ & $0$ & $-1$ & $0$ & $-1$  & $1$ \\\hline    
\hline
\end{tabular}
\end{center}
\caption{\small Low energy fermionic spectrum of the five stack 
string scale 
$SU(3)_C \otimes
SU(2)_L \otimes U(1)_a \otimes U(1)_b \otimes U(1)_c 
\otimes U(1)_d \otimes U(1)_e $, type I D6-brane model together with its
$U(1)$ charges. Note that at low energies only the SM gauge group 
$SU(3) \otimes SU(2)_L \otimes U(1)_Y$ survives.
\label{spectrum8}}
\end{table}
a)
There are various gauged low energy symmetries in the models. They are defined
in terms of the $U(1)$ symmetries $Q_a$, $Q_b$, $Q_c$, $Q_d$, $Q_e$. 
The baryon number B is equal to $Q_a = 3B$, 
the lepton number
is $L = Q_d + Q_e$ while   
$Q_a - 3 Q_d - 3 Q_e =  3 (B-L)$. Also note that $Q_c = 2 I_{3R}$, 
$I_{3R}$ being the third component of weak isospin. Also,
$3 (B-L)$ and $Q_c$ are free of triangle anomalies. The $U(1)_b$ symmetry
plays the role of a Peccei-Quinn symmetry in the sence of having mixed
SU(3) anomalies.
From the study of Green-Schwarz mechanism, see later chapters, we deduce 
that Baryon and Lepton number are unbroken gauged   
symmetries
and thus proton should be stable, while Majorana masses for right handed
neutrinos are not allowed.  
That means that mass terms for neutrinos should be of Dirac type.

The mixed anomalies $A_{ij}$ of the four surplus $U(1)$'s 
with the non-abelian gauge groups $SU(N_a)$ of the theory
cancel through a generalized GS mechanism \cite{iru, sa},
involving
 close string modes couplings to worldsheet gauge fields.
 Two combinations of the $U(1)$'s are anomalous and become massive, their
 orthogonal 
 non-anomalous combinations survive, combining to a single $U(1)$
 that remains massless, the hypercharge.
\newline
b) In order to cancel the appearance of exotic representations in the model
appearing from the $a a^{\star}$ sector, in antisymmetric and symmetric 
representations of the $U(N_a)$ group, we will impose the condition
\beq
 {\Pi}_{i=1}^3 m^i =\ 0. 
\label{req1}
\eeq
The solutions satisfying simultaneously the intersection constraints and the 
cancellation of the RR crosscap tadpole constraints
are parametric.
They are given in table (\ref{spectrum10}). 
The solutions represent
the most general solution of the RR tadpoles 
and they depend 
on five 
integer parameters $n_a^2$, $n_d^2$, $n_e^2$, $n_b^1$, $n_c^1$,
the phase parameters
$\epsilon = \pm 1$, ${\tilde \epsilon} = \pm 1$, and the NS-background
parameter
$\beta^i =\ 1-b^i$, which is associated to the presence of the 
NS B-field by $b^i =0,\ 1/2$. Note that the different solutions to the
tadpole constraints represent deformations of the D6 brane RR charges within
the same homology class.
In the rest of the paper we will be discussing, for simplicity the case with
$\epsilon =\ {\tilde  \epsilon} =\ 1$.  The multiparameter
tadpole solutions of table (2) represent deformations of the D6-brane
intersection spectrum of table (1), within the same homology class of the
factorizable three-cycles. 
By using the tadpole  
solutions of table (2) in (\ref{na1}) all tadpole equations but the 
first are automatically satisfied, the
\footnote{We have added an arbitrary number
of $N_D$ branes which don't contribute to the rest of the tadpoles and
intersection number constraints.}
latter yielding \footnote{We have set for
simplicity ${\tilde \epsilon } =1$.}:
\begin{table}[htb]\footnotesize
\renewcommand{\arraystretch}{3}
\begin{center}
\begin{tabular}{||c||c|c|c||}
\hline
\hline
$N_i$ & $(n_i^1, m_i^1)$ & $(n_i^2, m_i^2)$ & $(n_i^3, m_i^3)$\\
\hline\hline
 $N_a=3$ & $(1/\beta^1, 0)$  &
$(n_a^2,  \epsilon \b^2)$ & $(3, {\tilde \epsilon}/2)$  \\
\hline
$N_b=2$  & $(n_b^1, -\epsilon \b^1)$ & $(1/\beta^2, 0)$ &
$({\tilde \epsilon}, 1/2)$ \\
\hline
$N_c=1$ & $(n_c^1, \epsilon \b^1)$ &
$(1/\beta^2, 0)$  & $(0, 1)$ \\    
\hline
$N_d=1$ & $(1/\beta^1, 0)$ &  $(n_d^2,  2 \epsilon \b^2)$  
  & $(1, -{\tilde \epsilon}/2)$  \\\hline
$N_e = 1$ & $(1/\beta^1, 0)$ &  $(n_e^2,   \epsilon \b^2)$  
  & $(1, -{\tilde \epsilon}/2)$  \\
\hline
\end{tabular}
\end{center}
\caption{\small Tadpole solutions of 
D6-branes wrapping numbers giving rise to the 
standard model gauge group and spectrum at low energies.
The solutions depend 
on five integer parameters, 
$n_a^2$, $n_d^2$, $n_e^2$, $n_b^1$, $n_c^1$,
the NS-background $\beta^i$ and
the phase parameters $\epsilon = \pm 1$, ${\tilde \epsilon} = \pm 1$.
\label{spectrum10}}          
\end{table}
\beq
\frac{9 n_a^2}{ \b^1} +\ 2 \frac{n_b^1}{ \b^2} +\
\frac{n_d^2}{ \b^1} +\ \frac{n_e^2}{ \b^1} +\
N_D \frac{2}{\b^1 \b^2} =\ 16.
\label{ena11}
\eeq
Note that we had added the presence of extra $N_D$ branes 
(hidden branes). Their contribution to the RR tadpole conditions is best 
described by placing them in the three-factorizable cycle 
\beq 
N_D (1/\b^1, 0)(1/\b^2, 0)(2, m_D^3)
\label{sda12}
\eeq
and we have set $ m_D^3 =0$. 
To see clearly the cancellation of tadpoles, we have to choose a
consistent numerical set of wrappings, e.g.
\beq
 n_a^2 =1,\;n_b^1 =1,\;n_c^1 \in Z,\;n_d^2 =-1,
\;n_e =1,\;\b^1 =1/2,
\;\b^2 =1, \epsilon =1.
\label{numero1}
\eeq
With the above choices, all tadpole conditions but the first are 
satisfied, the latter is satisfied when we add
one ${\bar D}6$ brane, e.g. $N_D =\ -1$. 
Thus the tadpole structure \footnote{Note that the parameter
$n_c^1$ should be defined such that its choice is consistent with a 
tilted tori, e.g. $n_c^1 = 1$. } becomes
\begin{center}
\beqa
N_a =3&(2, \ 0)(1, \  1)(3,  1/2) \nonumber\\
N_b =2&(1,  -1/2)(1, \ 0) (1, \ 1/2) \nonumber\\
N_c =1&(n_c^1, \ 1/2)(1,\ 0)(0,\ 1) \nonumber\\    
N_d =1&(2, \ 0)(-1, 2)(1, -1/2)\nonumber\\
N_e =1&(2,\ 0) (1,\ 1)(1, \  -1/2).  
\label{consist}
\eeqa
\end{center} 
Actually, the satisfaction of the tadpole conditions  
is independent of $n_c^1$. Thus, when all other parameters are fixed,
$n_c^1$ is a global parameter that can vary. However, finally it will be 
fixed in terms of the remaining parameters once we specify, the tadpole 
subclass that corresponds to the massless spectrum with the hypercharge
embedding of the standard model. \newline
Note that there are always choices of wrapping numbers
of wrapping numbers that satisfy the RR tadpole 
constraints without the need of adding extra parallel
branes, e.g. the
following choice satisfies all RR
tadpoles
\beq
n_a^2 =1,\;n_b^1 =3,\;n_c^1 \in Z,\;n_d^2 =-3,
\;n_e =-1,\;\b^1 =1/2,
\;\b^2 =1, \epsilon =1.
\label{numero2}
\eeq
with cycle wrapping numbers 
\begin{center}
\beqa
N_a =3&(2, \ 0)(1, \  1)(3,  1/2) \nonumber\\
N_b =2&(3,  -1/2)(1, \ 0) (1, \ 1/2) \nonumber\\
N_c =1&(n_c^1, \ 1/2)(1,\ 0)(0,\ 1) \nonumber\\    
N_d =1&(2, \ 0)(-3, 2)(1, \ -1/2)\nonumber\\
N_e =1&(2,\ 0) (-1,\ 1)(1, \  -1/2).  
\label{consist1}
\eeqa
\end{center} 
Another alternative choice, satisfied by all RR tadpoles will be
\beq
n_a^2 =1,\;n_b^1 =1,\;n_c^1 \in Z,\;n_d^2 =2,
\;n_e =1,\;\b^1 =1,
\;\b^2 =1/2, \epsilon =1.
\label{numero3}
\eeq
with cycle wrapping numbers 
\begin{center}
\beqa
N_a =3&(1, \ 0)   (1, \ 1/2)     (3,  1/2) \nonumber\\
N_b =2&(1,  -1)    (2, \ 0)      (1, \ 1/2) \nonumber\\
N_c =1&(n_c^1, \ 1/2)(2,\ 0)    (0,\ 1) \nonumber\\    
N_d =1&(1, \ 0)     (2, 1)      (1, \ -1/2)\nonumber\\
N_e =1&(1,\ 0)      (1,\ 1/2)     (1, \  -1/2).  
\label{consist2}
\eeqa
\end{center}   

f) the hypercharge operator in the model is defined as a linear combination
of the three generators of the $SU(3)$, $U(1)_c$, $U(1)_d$, $U(1)_e$
gauge groups:
\beq
Y = \frac{1}{6}U(1)_{a}- \frac{1}{2} U(1)_c - \frac{1}{2} U(1)_d -
 \frac{1}{2} U(1)_e \;.
\label{hyper12}
\eeq

\section{Cancellation of U(1) Anomalies}

In general 
the mixed anomalies $A_{ij}$ of the four $U(1)$'s
with the non-Abelian gauge groups are given by
\beq
A_{ij}= \frac{1}{2}(I_{ij} - I_{i{j^{\star}}})N_i.
\label{ena9}
\eeq
Moreover, analyzing the mixed anomalies 
of the extra $U(1)$'s with the non-abelian gauge groups $SU(3)_c$, 
$SU(2)_b$, we can conclude that there are two anomaly free combinations
$Q_c$, $Q_a - 3 Q_d - 3 Q_e$.
Also, note that the gravitational anomalies cancel since D6-branes never 
intersect O6-planes.
In the orientifolded type I torus models gauge anomaly 
cancellation \cite{iru} \cite{iru1} is guaranteed through a 
generalized GS
mechanism \cite{louis2} that uses the 10-dimensional RR gauge fields
$C_2$ and $C_6$ and gives at four dimensions
the following couplings to gauge fields 
 \beqa
N_a m_a^1 m_a^2 m_a^3 \int_{M_4} B_2^o \wedge F_a &;& n_b^1 n_b^2 n_b^3
 \int_{M_4}
C^o \wedge F_b\wedge F_b,\\
N_a  n^J n^K m^I \int_{M_4}B_2^I\wedge F_a&;&n_b^I m_b^J m_b^K \int_{M_4}
C^I \wedge F_b\wedge F_b\;,
\label{ena66}
\eeqa
where
$C_2\equiv B_2^o$ and $B_2^I \equiv \int_{(T^2)^J \times (T^2)^K} C_6 $
with $I=1, 2, 3$ and $I \neq J \neq  K $. Notice the four dimensional duals
of $B_2^o,\ B_2^I$ :
\beqa
C^o \equiv \int_{(T^2)^1 \times (T^2)^2 \times (T^2)^3} C_6&;C^I \equiv
\int_{(T^2)^I} C_2, 
\label{ena7}
\eeqa
where $dC^o =-{\star} dB_2^o,\; dC^I=-{\star} dB_2^I$.

The triangle anomalies (\ref{ena9}) cancel from the existence of the
string amplitude involved in the GS mechanism \cite{sa} in four 
dimensions \cite{iru}. 
The latter amplitude, where the $U(1)_a$ gauge field couples to one
of the propagating
$B_2$ fields, coupled to dual scalars, that couple in turn to
two $SU(N)$ gauge bosons, is 
proportional \cite{louis2} to
\beq
-N_a  m^1_a m^2_a m^3_a n^1_b n^2_b n^3_b -
N_a \sum_I n^I_a n^J_a n^K_b m^I_a m^J_b m^K_b\; ,
I \neq J, K 
\label{ena8}
\eeq

Taking into account the phenomenological requirements of eqn.
(\ref{req1}) the RR couplings $B_2^I$ of (\ref{ena66}) then 
appear into three terms \footnote{For convenience we have included the
dependence on $\epsilon , {\tilde \epsilon}$ parameters.}:
\beqa
B_2^1 \wedge \left( \frac{- 2 \epsilon {\tilde \epsilon}    \b^1 }{\b^2 } 
\right)F^b,&\nonumber\\
B_2^2 \wedge \left(\frac{\epsilon \b^2}{\b^1}  
\right)(9F^a + 2   F^d+  F^e),&\nonumber\\
B_2^3  \wedge \left( \frac{3 {\tilde \epsilon} n_a^2}{2\b^1} F^a +     
\frac{n_b^1}{\b^2}F^b  + \frac{n_c^1}{\b^2} F^c -
\frac{{\tilde \epsilon} n_d^2}{2\b^1} F^d
-\frac{{\tilde \epsilon} n_e^2}{2 \b^1}F^e \right).&
\label{rr1}
\eeqa
At this point we should list 
the couplings of the dual scalars $C^I$ of $B_2^I$ that required to cancel
the mixed anomalies of the five $U(1)$'s with the 
non-abelian gauge groups $SU(N_a)$. They are given by
\beqa
C^1 \wedge [  \frac{\epsilon {\tilde \epsilon}\b^2 }{2 \b^1} (F^a \wedge 
F^a) - \frac{\epsilon {\tilde \epsilon} \b^2 }{ \b^1} (F^d \wedge F^d) -
\frac{ \epsilon  {\tilde \epsilon} \b^2 }{2 \b^1}
 F^e \wedge F^e )], &\nonumber\\
C^2 \wedge [ \frac{- \epsilon \b^1 }{2 \b^2 } (F^b \wedge 
F^b) +  \frac{ \epsilon \b^1 }{ \b^2 }(F^c \wedge F^c)    ],
&\nonumber\\
C^o \wedge \left(  \frac{3n_a^2}{\b^1}(F^a \wedge 
F^a)  +     \frac{{\tilde \epsilon}n_b^1}{ \b^2}(F^b \wedge 
F^b)  + \frac{n_d^2}{ \b^1}(F^d \wedge 
F^d)  + \frac{n_e^2}{ \b^1}(F^e \wedge F^e) \right),&
\label{rr2}
\eeqa

Notice that the RR scalar $B_2^0$ does not couple to any 
field $F^i$ as we have imposed the condition (\ref{req1}) 
which excludes the appearance of any exotic 
matter.\newline
Looking at (\ref{rr1}) we can conclude that
there are two anomalous $U(1)$'s that become massive through their 
couplings to the RR fields. They are the model independent fields,
 $U(1)_b $ and the combination $9U(1)_a +
 2U(1)_d + U(1)_e $,
which become massive through their couplings to the 
RR 2-form fields $B_2^1, B_2^2$ respectively.
In addition, there is
a model dependent, non-anomalous and massive $U(1)$ field
coupled to $B_2^3$ RR field.
That means that the 
 two massless and anomaly free combinations are 
$U(1)_c$ and $U(1)_a - 3U(1)_d -3 U(1)_e$.
Also, note that the 
 mixed anomalies $A_{ij}$ are cancelled 
by the GS mechanism set by the couplings (\ref{rr1}, \ref{rr2}).
\newline
The question we want to address at this point is how we can,
from the general
class 
of models, associated with the generic SM's of tables (1)
and (2), pick up the subclass
that corresponds to 
the ones associated with just the observable SM at low energies.
Clearly, for this to
happen we have to identify the subclass of tadpole solutions of table  
(\ref{spectrum10}) that corresponds to the hypercharge assignment
(\ref{hyper12}) of the
standard model \footnote{At this point, we recall an argument that have 
appeared in \cite{louis2}.} spectrum.
 
In general, the generalized Green-Schwarz mechanism
that cancels non-abelian anomalies of the $U(1)$'s to the 
non-abelian gauge fields involves couplings of closed string modes 
to the $U(1)$ field strengths \footnote{In addition,
to the couplings of the Poincare dual scalars $\eta_a$ of the fields $B_a$,
\beq
\sum_a g_a^k \ \eta_a \ tr(F^k \wedge F^k).  
\label{gree1}
\eeq
}
in the form
\beq
\sum_a \ f_a^i \  B_a\wedge tr(F_i).
\eeq
Effectively, the mixture of couplings in the form
\beq
A_{i \ k} +  \sum_a \ f_a^i \ g_a^k = \ 0 
\eeq
cancels the all non-abelian $U(1)$ gauge anomalies.
That means that if we want to keep some $U(1)$ massless we have to 
keep it decoupled from some closed string mode couplings that can make 
it massive, that is
 \beq
\sum_a \ ( \frac{1}{6} f_a^{\alpha} -\ \frac{1}{2}f_a^{c} -\ 
\frac{1}{2}f_a^{d} -\  \frac{1}{2}f_a^{e}  ) = \ 0\ .
\eeq
In conclusion,
the combination of the $U(1)$'s which remains light at low 
energies, 
is 
\beq
(3 n_a^2 + 3n_d^2 + 3 n_e^2) \neq 0,\;\ Q^l = n_c^1 (Q_a -3 Q_d -3 Q_e)
-\frac{3 {\tilde \epsilon}\b^2 ( n_a^2 + n_d^2 + n_e^2)}{2 \b^1} Q_c   .
\label{hyper}
\eeq
The subclass of tadpole solutions of (\ref{hyper}) having the SM hypercharge
assignment at low energies is exactly the one which is proportional to
(\ref{hyper12}). It
satisfies the condition, 
\beq
n_c^1 = \ \frac{{\tilde \epsilon}\b^2}{2\b^1}( n_a^2 +\  n_d^2 +\  n_e^2).
\label{mashyper}
\eeq

We note that there is one extra anomaly free, model
dependent $U(1)$
beyond the hypercharge
combination, and orthogonal to the latter, which 
is \footnote{ We note that alternatively, in an orthogonal basis, the three 
U(1)'s are coupled to $B_2^i$'s, the 4th U(1) is the hypercharge
(\ref{hyper}) and its condition (\ref{mashyper}), the fifth U(1) is 
$U(1)^{(5)} =(-\frac{3}{29}+ \frac{3}{28})F^a  -\frac{1}{29} F^d + 
\frac{1}{28} F^e$ , the latter surviving massless the Green-Schwarz mechanism
when $n_a^2 = -(28/9)n_d^2$. In this case, $U(1)^{(5)}$ should be broken
by the vev of $s\nu_R$ only (see section 5).}
\beq
Q^N =\ \frac{3{\tilde \epsilon}\b^2}{2 \b^1}(Q_a -\ 3Q_d -\ 3Q_e) + 19 n_c^1 Q_c .
\label{extra}
\eeq

Lets us summarize what we have found up to now. The tadpole solutions of
table (2), taking into account the condition (\ref{mashyper}), give at
low energies
classes of models that have the low
energy spectrum of the SM with the correct hypercharge assignments. At this
level the gauge group content of the model includes beyond $SU(3)
\otimes SU(2) \otimes U(1)_Y$ the additional $U(1)^N$ generator and
all SM particles gets charged under the additional $U(1)^N$ symmetry.
However, notice that $Q^N$ has a non-zero coupling to RR 
field $B_2^3$. That is it receives a mass of order of the
string scale $M_s$ and disappears from the
low energy spectrum. Hence at low energy only
the SM remains.

In the next sections we will see
that in the present five stack constructions
it is possible to use an additional mechanism
to break this extra
$U(1)$ symmetry, complementary to the one associated with
the RR fields. In involves the
usual mechanism of giving a vev
to a scalar field. In this case we will have to
require that the intersection where the right handed
neutrino is
localized, respects $N =\ 1$ supersymmetry.
In the latter case
the immediate
effect on obtaining just the SM at low energies will be
one additional
linear condition on the RR tadpole solutions of table (2).
We note that when $n_c^1 = 0$, it is possible
to have massless in the low energy spectrum both the $U(1)$ generators,
$Q_c$, and the B-L generator $(1/3)(Q_a -3 Q_d -3 Q_e)$ as long as
$n_c^1 = 0$, $n_a^2 = -n_d^2-n_e^2$.

\section{Electroweak Higgs and symmetry
breaking from open string tachyons}

The mechanism of electroweak symmetry breaking is a
well understood effect
at the level of gauge theories with or without
supersymmetry.
At the string theory level
the mechanism is believed to take place
either by using open string tachyonic modes between
parallel branes
or following a recent suggestion using brane 
recombination \cite{cim}.
In the former mechanism, the mass of the
Higgs field
receives contributions from the distance between
the branes $b$, $c$ ($b$, $c^{\star}$) which are parallel 
across the second
tori (see table 2). By varying the distance
between the branes, across the 2nd tori,
the Higgs mass could become tachyonic signalling
electroweak symmetry breaking.
In the latter mechanism, one of the two factorizable
b-branes, making
the $SU(2)$-stack, recombine with the single $U(1)$
c-brane into a single non-factorizable $j$-brane.
After the recombination the electroweak
symmetry is broken, a result better seen from
the new intersection numbers produced.
Note that the latter procedure is topological
and cannot be described using field theoretical methods.
During the recombination process instead of us working with
 our usual
wrapping numbers (\ref{na2}), one must work with
cycles associated with a non-factorizable $T^6$ torus. Also
one has to preserve the RR charge in homology before and
after the recombination process.
In this work, we will follow the former method.

\subsection{The angle structure}

We have up to now describe the appearance in the R-sector of open strings
of $I_{ab}$ massless chiral fermions
in the D-brane intersections that 
transform under bifundamental representations $N_a, {\bar N}_b$.
We should note that in backgrounds with intersecting 
branes, besides the actual
presence of massless fermions at each intersection, 
we have evident the presence of an equal number of
 massive scalars (MS), in the NS-sector, in exactly the same representations 
as the massless fermions \cite{luis1}. The mass of the MS
is of order of the string scale. In some cases,
it is possible that some of those MS may become 
tachyonic, triggering a potential that looks like the Higgs potential 
of the SM,
especially when their mass, that depends on the 
angles between the branes,
is such that is decreases the world volume of the 
3-cycles involved in the recombination process of joining the two
branes into a single one \cite{senn}.

The models examined in this work, are based on orientifolded six-tori
on type I
strings. In those configurations the bulk has ${\cal N} = 4$ SUSY.
Lets us now describe the open string sector of the model.
In order to describe the open string spectrum we introduce a four dimensional
twist \cite{tessera2, luis1} 
vector $\upsilon_{\theta}$, whose I-th entry
is given by  $\vartheta_{ij}$, with $\vartheta_{ij}$ the angle
between the branes $i$ and $j$-branes. After GSO projection
the states are labeled by a four dimensional twisted vector $r +
\upsilon_{\theta}$, where $\sum_I r^I =$odd
and $r_I  \in {\bf Z}, {\bf Z} {\bf + \frac{1}{2}}$
for NS, R sectors respectively. The Lorentz quantum numbers
are denoted by the last entry.
The mass operator for the states is provided by:

\beq
{\alpha^\prime} M^2_{ij} =\ \frac{Y^2}{4 \pi^2 \alpha^{\prime}} +\
N_{bos}(\vartheta) +\ \frac{(r + \upsilon)^2}{2} -\ \frac{1}{2} +\
E_{ij},
\label{mass}
\eeq
where $E_{ij}$ the contribution to the mass operator from bosonic
oscillators, and $N_{osc}(\vartheta)$ their number operator, with
\beq
E_{ij} =\  \sum_I \frac{1}{2}|\vartheta_I|(1 - |\vartheta_I|),
\label{bos}
\eeq
and $Y$ measures the minimum distance between branes for minimum
winding states.

If we
represent the twisted vector $r + \upsilon$,  by $(\vartheta_1,\vartheta_2,
\vartheta_3, 0)$, in the NS open string sector, the 
lowest lying states are given  \footnote{
we assumed $0\leq\vartheta_i\leq 1$ .} by:
{\small \beqa
\begin{array}{cc}
{\rm \bf State} \quad & \quad {\bf Mass} \\
(-1+\vartheta_1,\vartheta_2,\vartheta_3,0) & \alpha' M^2 =
\frac 12(-\vartheta_1+\vartheta_2+\vartheta_3) \\
(\vartheta_1,-1+\vartheta_2,\vartheta_3,0) & \alpha' M^2 =
\frac 12(\vartheta_1-\vartheta_2+\vartheta_3) \\
(\vartheta_1,\vartheta_2,-1+\vartheta_3,0) & \alpha' M^2 =
\frac 12(\vartheta_1+\vartheta_2-\vartheta_3) \\
(-1+\vartheta_1,-1+\vartheta_2,-1+\vartheta_3,0) & \alpha' M^2
= 1-\frac 12(\vartheta_1+\vartheta_2+\vartheta_3)
\label{tachdsix}
\end{array}
\eeqa}

The angles at the ten different intersections can be expressed
in terms of the tadpole solutions parameters.
Let us define the angles :
\beqa
\theta_1 \   = \ \frac{1}{\pi} tan^{-1}
\frac{\b^1 R_2^{(1)}}{n_b^1 R_1^{(1)}},  \
\theta_2 \  =   \  \frac{1}{\pi}
tan^{-1}\frac{\beta^2 R_2^{(2)}}{n_a^2 R_1^{(2)}}, \
\theta_3 \  = \  \frac{1}{\pi} tan^{-1}\frac{R_2^{(3)}}
{6 R_1^{(3)}},
 \nonumber \\
{\tilde {\theta_1}} \   = 
\ \frac{1}{\pi} tan^{-1}\frac{ \beta^1 R_2^{(1)}}{n_c^1 R_1^{(1)}},\;\
{\tilde { \theta_2}}  \ 
 = \ \frac{1}{\pi} tan^{-1} \frac{2 \beta^2 R_2^{(2)}}{n_d^2 R_1^{(2)}},   \;\
{\tilde {\theta_3 }} \  = \ \frac{1}{\pi} tan^{-1}
\frac{R_2^{(3)}}{2 R_1^{(3)}},\nonumber\\
 \ {\bar \theta}_2 \  =   \  \frac{1}{\pi} tan^{-1}
\frac{\beta^2 R_2^{(2)}}{n_e^2 R_1^{(2)}},    \ ;{\bar {\theta_3 }} \  = \
\frac{1}{\pi} tan^{-1}
\frac{R_2^{(3)}}{2 R_1^{(3)}},
\label{angulos}
\eeqa
where $R^{(i)}_{1,2}$ are the compactification radii
for the three $i=1,2,3$ tori, namely
projections 
of the radii 
 onto the $X^{(i)}_{1,2}$ directions when the NS flux B field,
$b^i$, is turned on and we have chosen for convenience $\epsilon =
{\tilde \epsilon} = \ 1$.

At each of the ten non-trivial intersections 
we have the 
presence of four states $t_i , i=1,\cdots, 4$, associated
to the states (\ref{tachdsix}).
 Hence we have a total of
forty different scalars in the model \footnote{ In figure one,
we can see the D6 branes angle setup in the present models.} .

\begin{figure}
\begin{center}
\centering
\epsfysize=10cm
\leavevmode
\epsfbox{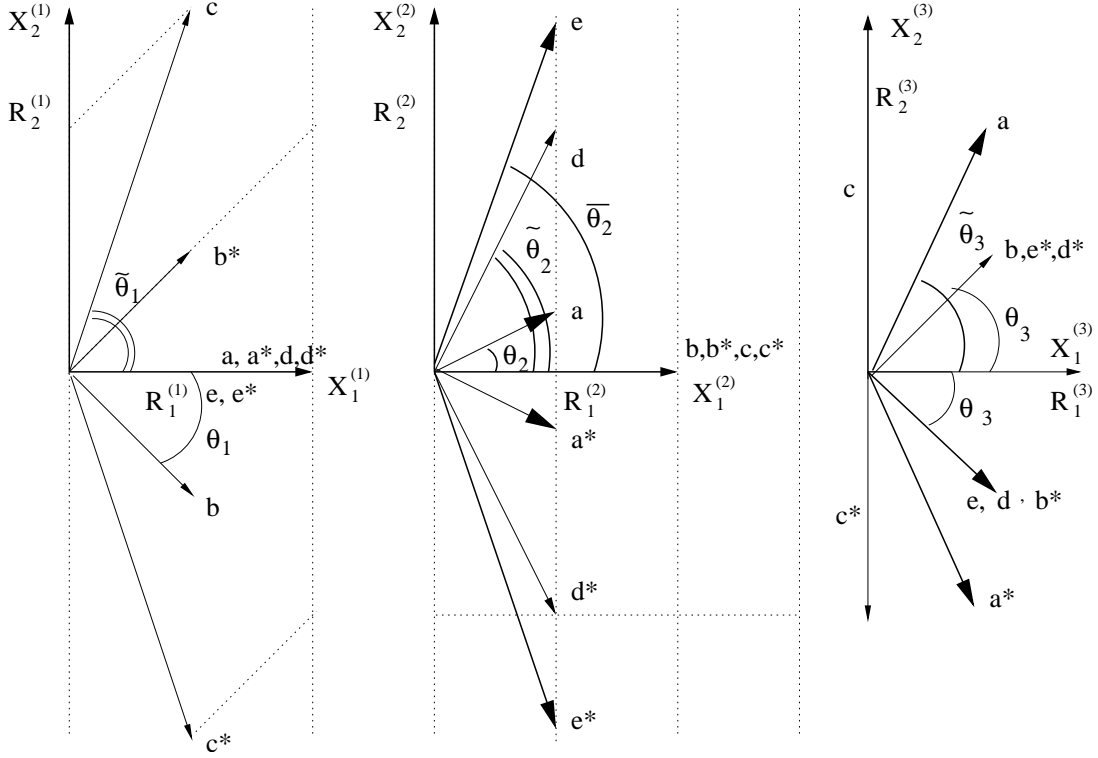}
\end{center}
\caption[]{\small
Assignment of angles between D6-branes on
the five stack type I model giving rise to the 
SM at low energies.
The angles between branes are shown on a product of 
$T^2 \times T^2 \times T^2$. We have chosen $ \b^1 =\b^2 =1$, $n_b^1, 
n_c^1, n_a^2, 
n_d^2 >0$, $\epsilon = {\tilde \epsilon} = 1$. }
\end{figure}

The following mass relations hold between the different
intersections of the classes of models :
\beqa
 m^2_{ab}(t_2)+m^2_{ab}(t_3) \ =\ m^2_{ab^{\star}}(t_2)+m^2_{ab^{\star}}(t_3)\
=\ m^2_{bd}(t_2)+m^2_{bd}(t_3),&
\nonumber \\
m^2_{ac}(t_2)+m^2_{ac}(t_3) \ =\ m^2_{ac^{\star}}(t_2)+m^2_{ac^{\star}}(t_3)
\ =\  m^2_{cd}(t_2)+m^2_{cd}(t_3),&\nonumber \\
m^2_{ce^{\star}}(t_1)+m^2_{ce^{\star}}(t_2) \ =\ m^2_{cd^{\star}}(t_1)+
m^2_{cd^{\star}}(t_2),&
\label{rela1}
\eeqa
or equivalently
\beqa
 m^2_{Q_L}(t_2)+m^2_{Q_L}(t_3) \ =\ m^2_{q_L}(t_2)+m^2_{q_L}(t_3)\
=\ m^2_{L}(t_2)+m^2_{L}(t_3),&
\nonumber\\
m^2_{U_R}(t_2)+m^2_{U_R}(t_3) \ =\ m^2_{D_R}(t_2)+m^2_{D_R}(t_3)
\ =\  m^2_{N_R}(t_2)+m^2_{N_R}(t_3),&\nonumber\\
m^2_{e_R}(t_1)+m^2_{e_R}(t_2) \ =\ m^2_{E_R}(t_1)+m^2_{E_R}(t_2),&
\label{rela11}
\eeqa

We note that in this work, we will not discuss the stability conditions
for absence of tachyonic scalars such that the D-brane configurations
discussed will be stable as this will be discussed elsewhere.
Similar conditions have been examined before in \cite{louis2, kokos1}.

\subsection{Tachyon Higgs mechanism in detail}

In this section, we will study the electroweak Higgs sector
of the models.
We note that below the string scale the massless spectrum of
the model is that of the SM with
all particles having the correct hypercharge assignments
but with the gauge symmetry being
$SU(3) \otimes SU(2) \otimes U(1) \otimes U(1)^N$. 
For the time being we will accept that the
additional $U(1)^N$ generator
breaks to a scale higher than
the scale of electroweak symmetry breaking. The latter issue will be
discussed in detail in the next section. Thus in the following we will
focus our attention to the Higgs sector of the theory.

In general, tachyonic scalars stretching between two 
different branes  can be used as Higgs scalars
as they can become non-tachyonic
by varying the distance between parallel branes.
This happens in the models under discussion as the complex scalars
$h^{\pm}$, $H^{\pm}$ get
localized between the $b$, $c$ and between $b$, $c^{\ast}$ branes
respectively and 
can be interpreted from the field theory point of view \cite{louis2} 
as Higgs fields which are responsible for the breaking the
electroweak symmetry.
We note that the intersection numbers of the $b,c$ and $b,c^{\star}$
branes across the six-dimensional torus vanish as a result of the fact
that the $b, c$ and $b, c^{\star}$ branes are
parallel across the second tori.
The electroweak Higgs fields, appearing as
$H_i$ (resp. $h_i$), $i=1,2$, in table (\ref{hig}), come from the
NS sector, from open
strings stretching between the
parallel $b$, $c^{\star}$ (resp. $c$)
branes along the
second tori, and from open strings stretching between intersecting
branes along
the first and third tori.

Initially, the Higgses of table (\ref{hig}),
are part of the massive spectrum of fields
localized in the intersections $bc, bc^{\star}$. However, we emphasize that
the Higgses $H_i$, $h_i$ become massless by varying the distance along the
second tori between the $b, c^{\star}$, $b, c$ branes respectively.
In fact, a similar set of Higgs fields appear in the four stack
models of \cite{louis2}, but obviously
with different geometrical data.
We should note that the representations of Higgs fields $H_i$,  $h_i$ is the
maximum allowed by quantization. Their number is model dependent.

\begin{table} [htb] \footnotesize
\renewcommand{\arraystretch}{1}
\begin{center}
\begin{tabular}{||c|c||c|c|c||}
\hline
\hline
Intersection & EW breaking Higgs & $Q_b$ & $Q_c$ &Y \\
\hline\hline
$b c$ & $h_1$  & $1$ & $-1$&1/2\\
\hline
$b c$  & $h_2$  & $-1$ & $1$&-1/2 \\\hline\hline
$b c^{\star}$ & $H_1$  & $-1$ & $-1$& 1/2\\ \hline
$b c^{\star}$  & $H_2$  & $1$ & $1$ & -1/2\\
\hline
\hline
\end{tabular}
\end{center}
\caption{\small Higgs fields responsible for
electroweak symmetry breaking.
\label{hig}}
\end{table}

The number of complex scalar doublets present in the models
is equal to the non-zero
intersection number product between the $bc$, $bc^{\star}$
branes in the
first and third complex planes. Thus
\beqa
n_{H^{\pm}} =\  I_{bc^{\star}}\ =\ |\epsilon \b_1(n_b^1 - n_c^1)|, &
n_{h^{\pm}} =\  I_{bc}\ =\ |\epsilon \b_1(n_b^1 + n_c^1)| .
\label{inter1}
\eeqa

The precise geometrical data for the scalar doublets are 
{\small \beqa
\begin{array}{cc}
{\rm \bf State} \quad & \quad {\bf Mass^2} \\
(-1+\vartheta_1, 0, \vartheta_3, 0) & \alpha' {\rm (Mass)}_Y^2 =
  { {Z_2}\over {4\pi ^2}}\ +\ \frac{1}{2}(\vartheta_3 - 
\vartheta_1) \\
(\vartheta_1, 0, -1+ \vartheta_3, , 0) & \alpha' {\rm (Mass)}_X^2 =
  { {Z_2}\over {4\pi ^2 }}\ +\ \frac{1}{2}(\vartheta_1 - \vartheta_3) \\
\label{Higgsmasses}
\end{array}
\eeqa}
where  $X = \{ H_{bc^{\star}}^{+}, h_{bc}^{+}  \}$,
$Y = \{ H_{bc^{\star}}^{-}, h_{bc}^{-}  \}$   and  $Z_2$ is the
distance$^2$ in transverse space along the
second torus,
$\vartheta_1$, $\vartheta_3$ are the (relative) angles 
between the $b$-, $c^{\star}$ (for $H^{\pm}$ )  (or $b$, $c$ for $h^{\pm}$)
branes in the
first and third complex planes.  

Also we note  
the presence of two "Higgsino masses" at each of the
$bc$ or $bc^{\star}$ intersections, with the same quantum numbers 
and representations as the
Higgs fields and masses    
corresponding to
{\small \beqa
\begin{array}{cc}
{\rm \bf State} \quad & \quad {\bf Mass^2} \\
(-1/2+\vartheta_1, \mp 1/2 , -1/2+\vartheta_3, \pm 1/2 ) &
{\rm (Mass)}^2 =
  { {Z_2}\over {4\pi ^2\alpha'}}.\  \\
\label{Higgsinomasses} 
\end{array}
\eeqa}
We note that in this picture while the Higgs fields can be made
massless by varying the distance between the branes,
the Higginos are not massless and are part of 
the $N =\ 2$ massive spectrum  accompanying the ``massless'' Higgs fields
at the intersections $bc$, $bc^{\star}$.

As we noted the presence of scalar doublets $H^{\pm}, h^{\pm}$, can be
seen as coming from the field theory mass matrix

\beq
(H_1^* \ H_2) 
\left(
\bf {M^2}
\right)
\left(
\begin{array}{c}
H_1 \\ H_2^*
\end{array}
\right)
+(h_1^* \ h_2)
\left(
\bf {m^2}
\right)
\left(
\begin{array}{c}
h_1 \\ h_2^*
\end{array}
\right) + h.c.
\eeq
where
\beqa
{\bf M^2}=M_s^2
\left(
\begin{array}{cc}
Z_2^{(bc^*)}&
\frac{1}{2}|\vartheta_1^{(bc^*)}-\vartheta_3^{(bc^*)}|  \\
\frac{1}{2}|\vartheta_1^{(bc^*)}-\vartheta_3^{(bc^*)}| &
Z_2^{(bc^*)}\\
\end{array}
\right),\\ \ {\bf m^2}=M_s^2
\left(
\begin{array}{cc}
Z_2^{(bc)}& \frac{1}{2}|\vartheta_1^{(bc)}-\vartheta_3^{(bc)}|  \\
\frac{1}{2}|\vartheta_1^{(bc)}-\vartheta_3^{(bc)}| & Z_2^{(bc)}
 \end{array} \right)
\eeqa
\vspace{1cm}
The fields $H_i$ and $h_i$ are thus defined as
\beq
H^{\pm}={1\over2}(H_1^*\pm H_2); \ h^{\pm}={1\over2}(h_1^*\pm h_2)  \ .
\eeq

As a result the effective potential which 
corresponds to the spectrum of Higgs scalars is given by
\beqa
V_{Higgs}\ =\ m_H^2 (|H_1|^2+|H_2|^2)\
+\ m_h^2 (|h_1|^2+|h_2|^2)\  \nonumber\\
+\ m_B^2 H_1H_2\
+\ m_B^2 h_1h_2\
+\ h.c.,
\label{Higgspot}
\eeqa
where
\beqa
 {m_h}^2 \ =\ { {Z_2^{(bc)}}\over {4\pi ^2\alpha '}}\ & ; & \ 
{m_H}^2 \ =\ {{Z_2^{(bc^*)}}\over {4\pi ^2\alpha '}}\nonumber \\
m_b^2\ =\ \frac{1}{2\alpha '}|\vartheta_1^{(bc)}-\vartheta_3^{(bc)}| \ & ;&
m_B^2\ =\ \frac{1}{2\alpha '}|\vartheta_1^{(bc^*)}-\vartheta_3^{(bc^*)}|
\label{masilla}
\eeqa

We note that the $Z_2$ is a free parameter, a moduli,  
and can become very small 
in relation to the Planck scale.
However, the $m_B^2$ mass can be expressed in terms of the scalar masses
of the particles present at the different intersections.
Going one step further, we can express the ``angle'' part of the Higgs masses
in terms of the angles defined in (\ref{angulos}) and in figure 1.
Explicitly, we find \footnote{We have chosen a configuration
with $\epsilon = {\tilde \epsilon }= 1$,  $n_b, n_c, n_d, n_e > 0 $.}  :
\beqa
m_B^2\ =\ \frac{1}{2\alpha '}|{\tilde \vartheta}_1 - \vartheta_1| \ & ;&
m_b^2\ =\ \frac{1}{2\alpha '}|{\tilde \vartheta}_1 + \vartheta_1 + \vartheta_3
-\frac{1}{2}|\nonumber\\
m_h^2\ =\   \frac{1}{2\alpha '}(\chi_b -\chi_c)^2 &;& m_H^2\ =\ 
\frac{1}{2\alpha '}(\chi_b + \chi_c)^2,
\label{analy1} 
\eeqa 
where $\chi_b$, $\chi_c$ the distances from the orientifold plane of the 
branes b, c.
 Making use of the scalar mass relations at the intersections of the model
we can reexpress the mass relations (\ref{analy1}), in terms of (\ref{rela1}).
The values of $m_B^2$, $m_b^2$ are given in appendix A.

\section{SUSY at intersections and intermediate scale }

In the present classes of models the $U(1)$ symmetry
gets broken and the associated
gauge boson receives a mass, as there is a non-zero
coupling of (\ref{extra}) to RR two form field $B_2^3$.
However, for special values of the
RR tadpole solution
parameters is is possible that the usual mechanism,
of giving a vev to a scalar, contributes additionally
to the mass of the massive gauge boson associated with
(\ref{extra}).

Thus 
our aim in this section is to provide us with a
complementary mechanism
to break the additional generator $U(1)^N$. That may happen by 
demanding that the sector $ce$ preserves $N=1$ SUSY. 
That will have as an effect the appearance of $I_{ce}$ massless scalars
in the intersection with the same quantum numbers as the massless
$I_{ce}$ fermions. Because $I_{ce}=1$, and the massless fermion localized
in the intersection is $\nu_R$, the massive partner
of $\nu_R$ which become
massless will be a $s\nu_R$.
Consequently, by giving a vev to $s\nu_R$, the $s\nu_R$ gets charged 
and thus breaks $U(1)^N$, leaving only the SM gauge group
$SU(3) \otimes SU(2) \otimes U(1)_Y$ at low energies.
Lets us describe the procedure in more detail.

We want the particles localized on the intersection $ce$ to respect
some amount of SUSY, in our case $N =\ 1$.
 That means
that the relative angle between branes $c$, $e$, should obey the SUSY 
preserving condition
\beq    
\pm {\tilde \theta}_1 \pm {\bar \theta}_2 \pm (\frac{\pi}{2} + \theta_3)
=\ 0
\label{pre1}
\eeq

In this case,  a massless scalar field appear in the intersection $ce$, the 
superpartner of the $\nu_R$, the $s\nu_R$ field. It is charged 
under the additional 
$U(1)^N$ symmetry, thus breaks $U(1)^N$ by receiving a vev. In this case
the surviving gauge symmetry is of SM.
The scale of the 
additional breaking, $M_N$, 
 is set from the vev of $s\nu_R$ and in principle, can be anywhere between
$M_Z$ and the string scale.

The following choice :

\beq
tan^{-1}{\frac{\b^1 U^{(1)}}{n_c^1}} + tan^{-1}\frac{\b^2 U^{(2)}}{n_e^2}
-  tan^{-1}(\frac{U^{(3)}}{6}) -\frac{\pi}{2} =\ 0, 
\label{sdad12}
\eeq
with
\beq
n_e^2 = 0,\ \;\;\frac{\b^1 U^{(1)} }{n_c^1} =\ \frac{U^{(3)}}{6},\;\;\
\alpha =\ \frac{\b^1 U^{(1)} }{n_c^1},\;\;
U^{(i)} =\ \frac{R_2^{(i)} }{ R_1^{(i)}}.
\label{solu1}
\eeq
solves (\ref{pre1}).
In particular,
\beq
n_e^2=0 \ \Rightarrow \ \b^2=1
\label{condu}
\eeq
thus the second tori is not
tilted. The angle content of the 
branes, $c$, $e$, when the gauge symmetry breaks to the SM is given in
table (\ref{higg}).
\begin{table} [htb] \footnotesize
\renewcommand{\arraystretch}{1}
\begin{center}
\begin{tabular}{||c|c||c|c||}
\hline
\hline
Brane & $\theta_a^1$ & $\theta_a^2$ & $\theta_a^3$ \\
\hline\hline
$c$ & $tan^{-1}\alpha$  & $0$ & $\frac{\pi}{2}$\\
\hline
$e$  & $0$  & $\frac{\pi}{2}$ & $tan^{-1}(\alpha)$\\\hline\hline
\end{tabular}
\end{center}
\caption{\small Angle content for branes participating in the 
gauge symmetry breaking to the SM. Imposing ${\cal N} =\ 1$ SUSY on the
open sector $ce$ breaks the surplus $U(1)^N$ by a vev of $s\nu_R$.
\label{higg}}
\end{table}

Summarizing
the SM exists
between  $M_Z$ and the mass scale $M_N$ of the additional
$U(1)^N$. A set of SM wrappings exists only if we consider the
{\em hypercharge} (\ref{mashyper}) and the {\em gauge symmetry breaking}
condition (\ref{solu1}) when defining numerically the tadpole
solutions of table (2).
Taking into account both conditions a consistent set, for the observable SM
to exist,
wrapping numbers
is given by
\beq
n_e^2 =\ 0,\; \b^2=1,\;\b^1=1/2,\;n_b=-1,\;n_d^2=-1,\;n_a^2=2,\; n_c^1=1
\eeq
or
\begin{center}
\beqa
N_a=3&(2,  0)  (2,   1)  (3,   1/2) \nonumber\\
N_b=2&(-1,  -1/2)  (1,  0) (1,  1/2) \nonumber\\
N_c=1&(1,  1/2) (1, 0)  (0, 1)\nonumber\\    
N_d=1&(2,  0) (-1, 2) (1, -1/2) \nonumber\\
N_e=1&(2, 0)  (0, 1)  (1, -1/2). 
\eeqa
\end{center}
It satisfies all tadpole conditions but the first, the latter is satisfied
with the addition of four ${\bar D}_6$ located at 
$(2,0)(1,0)(2,0)$.

The number of electroweak Higgs present in the model can be investigated
further. The most interesting cases that have a minimal Higgs content
 follow :

\begin{itemize}

\item {\em The Higgs system of  MSSM}

For (\ref{inter1}), we can see that the minimal set of Higgs in
the models is
obtained for either $n_H = 0$, $n_h = 1$, or $n_h =1$, $n_H = 0$.
The two cases, are studied in table (\ref{tablo}). We found two families of
models that depend on a single integer $n_d^2$. We also list the number 
of necessary $N_D$ branes
 required to cancel the first tadpole condition.
 We have taken into account the conditions
 (\ref{mashyper}), (\ref{condu})
 necessary to obtain the observable SM at low energies.

The case $n_H = 1$, $n_h = 0$  appears to be the most interesting as this
appears to give a plausible explanation for the existence of small 
and different
neutrino masses
to the different generations. These issues are examined in more detail in
the next section.

\begin{table}[htb] \footnotesize
\renewcommand{\arraystretch}{1.5}
\begin{center}
\begin{tabular}{||c|c|c|c|c|c|c||}
\hline\hline
Higgs fields & $\b_1$ & $\b_2$ & $n_b$ & $n_c$
& $n_a^2$ & $N_D$ \\\hline
$n_H=1,\  n_h =0$ & $1/2$ & $1$ & $1$ & $-1$ &$-1 -\ n_d^2 $ & $8 + 4n_d^2$ \\\hline
$n_H=1, \ n_h =0$ &  $1/2$ & $1$ & $-1$ & $1$ & $1-\ n_d^2$ & $4n_d^2 $  \\\hline\hline
$n_H=0, \ n_h =1$ & $1/2$ & $1$ & $1$ & $1$ & $1-n_d^2$ & $-1+4n_d^2$ \\\hline    
$n_h=0, \ n_h =1$ &   $1/2$  &  $1$ &  $-1$ & $-1$ & $-1-n_d^2$ & $9+4n_d^2$ \\\hline\hline    
$n_H=1, \ n_h =1$ &   $1$  &  $1$ &  $1$ & $0$ & $-n_d^2$ & $7+4n_d^2$  \\\hline    
$n_H=1, \ n_h =1 $ &   $1$  &  $1$ &  $-1$ & $0$ & $-n_d^2$ & $9+4n_d^2$  \\\hline    
$n_H=1, \ n_h =1 $ &  $1$ &  $1$ & $0$ & $1$ & $2 -n_d^2$ & $-1+4n_d^2$   \\\hline
  $n_H=1, \ n_h =1 $ &   $1$  &  $1$ &   $0$ & $-1$ & $-2 -n_d^2$ & $16+4n_d^2$ \\\hline    
\hline
\end{tabular}
\end{center}
\caption{\small Families of models with with minimal Higgs structure. They
depend on a single integer, $n_d$. The surplus {\em gauge symmetry breaking}
condition (\ref{condu}) has been taken into account.
\label{tablo}}
\end{table}

\item {\em  Double Higgs system}
The next to minimal set of Higgses is obtained when $n_H=1, \ n_h =1$. In this 
case, quarks and leptons get their mass from the start. 
\end{itemize}

\section{Neutrino couplings and masses}

The Yukawa couplings in this model follow the usual pattern that
appears
in intersecting branes \cite{luis1}. 
The couplings
 between the two fermion states
$F_L^i$, ${\bar F}_R^j$ and the Higgs fields $H^k$,  arise from the 
stretching of the worldsheet between the three D6-branes which cross 
at those intersections.
For a six dimensional torus they can take the following form in the
leading order \cite{luis1},
\beq
Y^{klm}=e^{- {\tilde A}_{klm}},
\label{yuk1}
\eeq
where ${\tilde A}_{klm}$ is the worldsheet area connecting the three vertices
in the six dimensional space.
The areas of each of the two dimensional torus involved in
this interaction is typically of order one in string units.
For the models discussed in table (1), 
the Yukawa interactions for the chiral spectrum of the SM's yield:
\beqa
Y_j^U Q_L U_R^j h_1 + Y_j^D Q_L D_R^j H_2 &+ \nonumber\\
Y_{ij}^u q_L^i U_R^j H_1 + Y_{ij}^d q_L^i D_R^j h_2 +&\nonumber\\
Y_h^l \ l_L^h \ {\nu}_R^h \ h_1 +\ Y_h^e l_L^h e_R^h H_2 &+ \nonumber\\
Y_{ij}^N L^i N_R^j h_1 + Y_{ij}^E L_i E_R^j H_2 +\ h.c& \nonumber\\
\label{era1} 
\eeqa
where $i=1, 2$, $j=1,2,3$, $h=1$.

The nature of Yukawa couplings is such that the lepton and neutrino sector of 
the models distinguish between different generations, e.g. the "first" from 
the other two generations, as one generation
of neutrinos (resp. leptons)
is placed on a different intersection from the other two one's.
For example looking at the charged leptons of table (1) we see that
one generation of charged leptons $l_L$ gets localized on $be$-intersection,
while the other two generations $L$ get localized in the $bd$-intersection.

There are a number of scalar doublets in the model present that are
interpreted in terms of the low energy theory as Higgs doublets.
The most interesting ones were mentioned briefly in the last section.
There are two possibilities to be discussed.
The minimal case when $n_H = 1$, $n_h = 0$ and the next to minimal case,
  $n_H = 1$, $n_h = 1$.   

\begin{itemize}

\item {\em Minimal Higgs presence}
\newline
Without loss of generality we choose $n_H = 1$, $n_h = 0$, 
only the $H_1$, $H_2$ fields are present. The mechanism that will
give masses to the charged
lepton/quark
sector is similar to what happended in the four stack models
of \cite{louis2}.
At tree level two U-quarks and one D-quark as well the charge leptons gain
masses. Identifying the massive quarks with t, c, b, the
rest of the
quarks as well as neutrinos remain massless at tree level. The rest of the
quarks are expected to receive masses from strong interaction effects,
that create 
effective couplings
of the form $Q_L U^j_R H_1$, $q_L^iD_R^j H_2$ creating
masses for the
u-, d-, s-quarks of less than equal of $\Lambda_{QCD}$.
 Note that the
latter couplings are not
allowed at tree level since otherwise the $U(1)_b$ global symmetry
would have been violated.

The neutrino sector is slightly different from the four stack
counterparts \cite{louis2}  
of the SM's discussed in the present work.
Contrary to the models of \cite{louis2} where all neutrinos
come from the same
intersection, in this work
the SM's  
are build such that one generation of
neutrinos (and leptons) are placed at a different intersection from the
remaining two generations \footnote{The latter though are placed both at the same
intersection.}.
Thus the structure of Yukawa couplings in the neutrino sector
suggests that they is a distinction between the neutrinos of the
one (e.g. first)
generation and the other two one's. That could be used in 
principle to discuss neutrino mass textures in the context  
of recent results \footnote{see for example ref. \cite{gia}.} 
of SuperKamiokande \cite{super}, which suggest
that at least two generations of neutrinos may be massive.

Because Lepton number is a gauged symmetry, there are only Dirac masses
allowed. Thus a see-saw mechanism is excluded. The origin of
small neutrino masses originates from
dimension 6 operators in the form, breaking the $U(1)_b$ PQ like symmetry
through chiral symmetry breaking,
\beqa
\alpha^{\prime}(L N_R)\ (Q_LU_R)^{\star},\;\; \alpha^{\prime}(l \nu_R)
(q_L U_R).
\label{smnu}
\eeqa
Hence, the smallness of neutrino masses is related to the existence of the
dominant u-quark chiral condensate, $<u_R u_R>$. For
values \footnote{See for example  ref. \cite{QCD}.}
of the u-chiral
condensate, and assuming that all generations of neutrino species
receive the same value for u-condensate,
of order $(240 MeV)^3$, the neutrinos get a mass of order
\beq
\frac{<u_R u_L>}{M_s^2} =\ \frac{(240 MeV)^3}{M_s^2}
\label{order}
\eeq
Hence neutrino masses of order 0.1 - 10 eV are easily achieved in consistency
with LSND oscillation experiments \cite{LSND}.
In this case, a generation mixing among neutrino species is generated
from the leading order coupling behavior (\ref{yuk1}).

Alternatively, if we assume that the chiral condensate
generates the generation
mixing, receiving
different values for different neutrino species the neutrino masses
will depend weakly on the precise form of the couplings (\ref{yuk1}).

\item {\em Next to minimal Higgs presence} \newline
In this case, all couplings to quarks and leptons are realized from
the start.
All particles get a mass. The hierarchy of masses depend both on the
Higgs fields and the leading order Yukawa behavior (\ref{yuk1}).

\end{itemize}

\section{Conclusions and future directions}

In this work, we have presented the first examples
of string models, not based on a GUT group at
the string scale,
that have at low energies only the standard model
and are derived from five stacks of ( D6 ) branes at the string
scale \footnote{The first examples of models giving rise to the SM at low
energies  have been
considered in \cite{louis2}, using four stacks of branes and the same
type I orientifolded $T^6$ constructions.}. We note the following :

$\bullet$ Baryon number is a gauged symmetry and proton is stable.    
If proton was not stable in the models then we should have push the
string scale higher than $10^{16}$ GeV to suppress
dimension six operators that could potentially contribute to 
proton decay. However, in the present class of models, this is not 
necessary
as proton is stable.

$\bullet$ 
The additional $U(1)^N$ symmetry may break 
from its non-zero coupling to   
the RR two form fields.

However, as we have noted an 
additional mechanism is available for
certain values of the
tadpole and complex structure parameters, and thus
complementary to the Abelian gauge field mass
term generation induced
by their non-zero couplings to
the RR two form fields $B_2^i$, $i = 1, 2,3$.
 Crucial for this new mechanism, thus contributing to the breaking of the 
additional anomaly free
$U(1)$ generator and getting only the SM at low energies, was the
novel partial imposition of ${\cal N} =\ 1$ supersymmetry at only one
intersection, e.g.
$ce$. That had as an immediate effect to pull out from the massive
modes the superpartner
of $\nu_R$. 
It would be interesting to investigate in detail the symmetry
breaking patterns that follow from having a SUSY intersection, at an open
string sector, of the non-SUSY SM's examined in this work.

$\bullet$ We emphasize that the breaking of the $U(1)^N$
symmetry implies the
existence of an extra $Z^o$ boson above the electroweak scale.
Bounds on additional gauge bosons exist \cite{groom}
placing them in the 
range between 500-800 GeV. Thus we conclude that the string scale
should be at least equal to $M_N$ or higher.
Improved bounds of the string scale for the
present models
would require a generalization of the four stack D6 model \cite{louis2}
analysis of \cite{que}, for the masses of the
extra $U(1)$ gauge bosons made massive by the
Green-Schwarz mechanism.
 It would be interesting to extend the analysis
of \cite{que} to the present SM's that predict an
additional intermediate scale between $M_Z$ and $M_S$.

$\bullet$ A natural extension of the
five-stack D6-brane SM's of this work is to examine
how we can construct D6-brane models that respect some
supersymmetry at every intersection as was
detailed \footnote{see the first two references
of \cite{cim} for a similar construction for the 4-stack SM's
of \cite{louis2}.} recently.

$\bullet$ The models have vanishing RR tadpoles but some NSNS tadpoles remain,
leaving a open issue the full stability of the configurations. It is
then an
open question if the backgrounds can be cured using
Fischler-Susskind mechanism \cite{fi} in 
redefining the background \cite{nsns} as in \cite{rasu}.

Concluding this work, it is very interesting
that the present class of models predicts not only the existence
of a non-supersymmetric standard model at low energies but
in addition other classes of models predicting 
the unique existence of a SUSY partner of the right handed neutrino, the
$s\nu_R$.

\section{Acknowledgments}
I am grateful to D. Cremades, L. Ib\'a\~nez and 
A. Uranga for usuful discussions.

\newpage
\section{Appendix A}

In this appendix, we list the values of the mass parameters, of section
4,
involved in the mass of the set of four Higgses taking part in the 
process of electroweak symmetry breaking.  
As we remark in the main body of the paper, the quadratic parts of the 
Higgs mixing mass terms in the effective field theory potential are 
exactly calculable at tree level in the D6-brane models. 

\begin{center}
\beqa
m_B^2 =\frac{1}{2 } |m^2_{Q_L }(t_2) + m^2_{Q_L }(t_3) - 
m^2_{U_R}(t_2) - m^2_{U_R}(t_3)|\nonumber\\
  = \frac{1}{2}|  m^2_{Q_L}(t_2) + m^2_{Q_L}(t_3)  
   - m^2_{D_R}(t_2)-
   m^2_{D_R}(t_3)|\nonumber\\
  =  \frac{1}{2}|  m^2_{Q_L}(t_2) + m^2_{Q_L}(t_3)  
   - m^2_{N_R}(t_2)-
   m^2_{N_R}(t_3)|\nonumber\\
=  \frac{1}{2} |m^2_{q_L}(t_2) + m^2_{q_L}(t_3) - 
m^2_{U_R}(t_2) - m^2_{U_R}(t_3)|\nonumber\\
 =  \frac{1}{2}|  m^2_{q_L}(t_2) + m^2_{q_L}(t_3)  
   - m^2_{D_R}(t_2)-
   m^2_{D_R}(t_3)|\nonumber\\
     =  \frac{1}{2}|  m^2_{q_L}(t_2) + m^2_{q_L}(t_3)  
   - m^2_{N_R}(t_2)-
   m^2_{N_R}(t_3)|\nonumber\\
=  \frac{1}{2} |m^2_{L }(t_2) + m^2_{L}(t_3) - 
m^2_{U_R }(t_2) - m^2_{U_R }(t_3)|\nonumber\\
 =  \frac{1}{2}|  m^2_{L}(t_2) + m^2_{L }(t_3)  
   - m^2_{D_R}(t_2)-
   m^2_{D_R}(t_3)|\nonumber\\
     =  \frac{1}{2}|  m^2_{L}(t_2) + m^2_{L}(t_3)  
   - m^2_{N_R}(t_2)-
   m^2_{N_R}(t_3)|\nonumber\\
\eeqa
\end{center}
\begin{center}
\beqa
m_b^2 =\frac{1}{2 } |m^2_{Q_L }(t_2) + m^2_{Q_L }(t_3) +
m^2_{U_R}(t_2) + m^2_{U_R}(t_3)
-m^2_{e_R}(t_1) - m^2_{e_R}(t_2)|\nonumber\\
=\frac{1}{2} |m^2_{Q_L }(t_2) + m^2_{Q_L }(t_3) +
m^2_{U_R}(t_2) + m^2_{U_R}(t_3)
-m^2_{E_R}(t_1) - m^2_{E_R}(t_2)|\nonumber\\
  =  \frac{1}{2}|  m^2_{Q_L}(t_2) + m^2_{Q_L}(t_3)  
   + m^2_{N_R}(t_2)+
  m^2_{N_R}(t_3)-m^2_{e_R}(t_1) - m^2_{e_R}(t_2)|\nonumber\\
=  \frac{1}{2}|  m^2_{Q_L}(t_2) + m^2_{Q_L}(t_3)  
   + m^2_{N_R}(t_2)+
   m^2_{N_R}(t_3) -m^2_{E_R}(t_1) - m^2_{E_R}(t_2)  |\nonumber\\
=  \frac{1}{2} |m^2_{q_L}(t_2) + m^2_{q_L}(t_3) +
m^2_{U_R}(t_2) + m^2_{U_R}(t_3)-m^2_{e_R}(t_1) - m^2_{e_R}(t_2)|\nonumber\\
=  \frac{1}{2} |m^2_{q_L}(t_2) + m^2_{q_L}(t_3) +
m^2_{U_R}(t_2) + m^2_{U_R}(t_3) -m^2_{E_R}(t_1) - m^2_{E_R}(t_2) |\nonumber\\
 =  \frac{1}{2}|  m^2_{q_L}(t_2) + m^2_{q_L}(t_3)  
   + m^2_{D_R}(t_2)+
   m^2_{D_R}(t_3) -m^2_{e_R}(t_1) - m^2_{e_R}(t_2)   |\nonumber\\
=  \frac{1}{2}|  m^2_{q_L}(t_2) + m^2_{q_L}(t_3)  
   + m^2_{D_R}(t_2)+
   m^2_{D_R}(t_3) -m^2_{E_R}(t_1) - m^2_{E_R}(t_2)   |\nonumber\\
     =  \frac{1}{2}|  m^2_{q_L}(t_2) + m^2_{q_L}(t_3)  
   + m^2_{N_R}(t_2)+
   m^2_{N_R}(t_3) -m^2_{e_R}(t_1) - m^2_{e_R}(t_2)   |\nonumber\\
 =  \frac{1}{2}|  m^2_{q_L}(t_2) + m^2_{q_L}(t_3)  
   + m^2_{N_R}(t_2)+
   m^2_{N_R}(t_3)  -m^2_{E_R}(t_1) - m^2_{E_R}(t_2)   |\nonumber\\
=  \frac{1}{2} |m^2_{L }(t_2) + m^2_{L}(t_3) +
m^2_{U_R }(t_2) + m^2_{U_R }(t_3)  -m^2_{e_R}(t_1) - 
m^2_{e_R}(t_2)|\nonumber\\
=  \frac{1}{2} |m^2_{L }(t_2) + m^2_{L}(t_3) +
m^2_{U_R }(t_2) + m^2_{U_R }(t_3) -m^2_{E_R}(t_1) 
- m^2_{E_R}(t_2)|\nonumber\\
 =  \frac{1}{2}|  m^2_{L}(t_2) + m^2_{L }(t_3)  
   + m^2_{D_R}(t_2)+
   m^2_{D_R}(t_3)    -m^2_{e_R}(t_1) - 
m^2_{e_R}(t_2)  |\nonumber\\
 =  \frac{1}{2}|  m^2_{L}(t_2) + m^2_{L }(t_3)  
   + m^2_{D_R}(t_2)+
   m^2_{D_R}(t_3) -m^2_{E_R}(t_1) 
- m^2_{E_R}(t_2)  |\nonumber\\
     =  \frac{1}{2}|  m^2_{L}(t_2) + m^2_{L}(t_3)  
   + m^2_{N_R}(t_2)+
   m^2_{N_R}(t_3)   -m^2_{e_R}(t_1) - 
m^2_{e_R}(t_2)    |\nonumber\\
 =  \frac{1}{2}|  m^2_{L}(t_2) + m^2_{L}(t_3)  
   + m^2_{N_R}(t_2)+
   m^2_{N_R}(t_3)   -m^2_{E_R}(t_1) 
- m^2_{E_R}(t_2)  |\nonumber\\
\eeqa
\end{center}

\end{document}